\documentclass[12pt]{article}
\usepackage{graphicx}

\newcommand{\dev}{\mathrm{d}}
\newcommand{\el}{\vspace{1\baselineskip}\vspace{-\parskip}}

\newcommand\pubnumber{Article 24 in
eConf C1304143}
\newcommand\pubdate{\today}

\def\napoli{Center for Cosmology \& Particle Physics\\
Physics Department, New York University, New York, NY 10003, USA\\
Alexander von Humboldt Fellow\\
Max-Planck Institute for Extraterrestrial Physics (MPE)\\ Postfach 1312, 85741 Garching, GERMANY}

\def\Title#1{\begin{center} {\Large #1 } \end{center}}
\def\Author#1{\begin{center}{ \sc #1} \end{center}}
\def\Address#1{\begin{center}{ \it #1} \end{center}}

\newcommand\pubblock{\rightline{\begin{tabular}{l} \pubnumber\\
         \pubdate  \end{tabular}}}
\newenvironment{Abstract}{\begin{quotation}  }{\end{quotation}}
\newenvironment{Presented}{\begin{quotation} \begin{center} 
             PRESENTED AT\end{center}\bigskip 
      \begin{center}\begin{large}}{\end{large}\end{center} \end{quotation}}
\def\Acknowledgements{\bigskip  \bigskip \begin{center} \begin{large}
             \bf ACKNOWLEDGEMENTS \end{large}\end{center}}

\def\beq{\begin{equation}}
\def\eeq#1{\label{#1}\end{equation}}
\def\eeqn{\end{equation}}

\def\beqa{\begin{eqnarray}}
\def\eeqa#1{\label{#1}\end{eqnarray}}
\def\eeqan{\end{eqnarray}}

\let\bar=\overbar

\def\Dslash{\not{\hbox{\kern-4pt $D$}}}
\def\dslash{\not{\hbox{\kern-2pt $\del$}}}

\def\msb{{\bar{\ssstyle M \kern -1pt S}}}

\begin{document}
\begin{titlepage}
\pubblock

\vfill
\Title{Gamma-ray burst afterglow theory}
\vfill
\Author{ Hendrik van Eerten}
\Address{\napoli}
\vfill
\begin{Abstract}
It is by now fairly well established that gamma-ray burst afterglows result from initially relativistic outflows interacting with the medium surrounding the burster and emitting non-thermal radiation ranging from radio to X-rays. However, beyond that, many big and small questions remain about afterglows, with the accumulating amount of observational data at the various frequencies raising as many questions as they answer. In this review I highlight a number of current theoretical issues and how they fit or do not fit within our basic theoretical framework. In addition to theoretical progress I will also emphasize the increasing role and usefulness of numerical studies of afterglow blast waves and their radiation.
\end{Abstract}
\vfill
\begin{Presented}
GRB 2013, Huntsville Gamma - Ray Burst Symposium
14-18 April 2013 Nashville, Tennessee
\end{Presented}
\vfill
\end{titlepage}
\def\thefootnote{\fnsymbol{footnote}}
\setcounter{footnote}{0}

\section{Introduction}

Gamma-ray burst (GRB) afterglows occupy a unique position among the various high-energy astrophysical outflow phenomena. They are extremely relativistic blast waves with inferred Lorentz factors $\gamma$ that can be over several hundreds, far more than those typical for active galactic nuclei ($\gamma \sim 25$) or microquasars ($\gamma \sim 5$). They are transient events that occur only once per source. And they are relatively `clean', certainly when compared to the prompt emission, in that their outflows are (at least eventually) not dominated by complex large scale magnetic fields and in that their broadband emission from radio to X-rays is dominated by a single radiative process (synchrotron emission). 

This picture, of course, becomes more murky the earlier one looks and the closer to the prompt emission one gets, and the more one looks in detail at the peculiarities of any given afterglow dataset. But broadly speaking, the main conceptual issues with respect to afterglow blast waves are (1) the geometry and dynamics of the hydrodynamical outflow and the structure of its environment, (2) the microphysics of shock acceleration of electrons and the generation of fields at the shock front and (3) how the previous two lead to local emission that combines and leads to a global synchrotron-type spectrum that is observable at cosmological distances. Because the blast waves move with nearly the speed of light, the bookkeeping effort in step (3) depends sensitively on the evolution of the blast wave during the timespan in which simultaneously arriving radiation is emitted from various parts of the outflow.

In this review I focus mostly on the most basic afterglow model, where a collisionless shock wave interacts with a circumburst medium. This scenario was originally predicted in the context of the fireball model \cite{Rees1992} but is not unique to it. Even initially magnetically dominated outflows \cite{Spruit2001} or ballistic ejecta \cite{Dado2002} will eventually lead to a blast wave of swept-up material at further distance from the progenitor. Already in its simplest form, this basic model gives rise to a wide range of observational consequences and poses a number of computational challenges. The purpose of this review is to highlight these and to identify some limitations and aspects not emphasized elsewhere. I benefit from the fact that a number of important issues regarding GRB afterglows are already reviewed elsewhere in these proceedings, such as flares, energy injection (in the context of magnetars), afterglow polarization and short GRBs.

\section{Basic dynamics of a relativistic blast wave}

The most basic model for the afterglow dynamics is that of an initially relativistic explosion collimated with half-opening angle $\theta_0$ and with isotropic equivalent energy $E_{iso}$ adiabatically expanding in a cold homogeneous medium of density $\rho_{ext}$. Once the blast wave has reached a radius far greater than its initial radius, i.e. $r \gg r_0$, at time $t \gg t_0$ and the energy in the swept-up external mass greatly exceeds that in the initial mass of the ejecta (if any), the hydrodynamical equations for the blast wave will be functions only of $\theta_0$, $E_{iso}$, $\rho_{ext}$, speed of light $c$ and coordinates $r$, $\theta$, $t$ (assuming symmetry along $\phi$). Before the launch of Swift with its fast slewing capabilities, this was also effectively the only stage of the afterglow that was observed.

Instead of using $r$, $t$ and $\theta$, the fluid equations can be written in terms of dimensionless combinations $A \equiv r c  / t$, $B \equiv E_{iso} t^2 / (\rho_{ext} r^5)$, $\theta$. These variables are invariant under any transformation $E'_{iso} = \kappa E_{iso}$, $\rho'_{ext} = \lambda \rho_{ext}$, $r' = (\kappa / \lambda)^{1/3} r$, $t' = (\kappa / \lambda)^{1/3} t$ and from this straightforward dimensional analysis it follows that for a given initial opening angle $\theta_0$, the blast wave goes through exactly the same stages when explosion energy is increased (or circumburst density decreased), but at larger radii and later times. In a practical sense, this significantly reduces the parameter space for numerical simulations, to an extent that it can be fully covered and utilized for data analysis \cite{vanEerten:2011yn}.

In the ultrarelativistic stage at early times there is no causal contact yet along the different angles of the blast wave since the comoving speed of sound has a finite relativistic upper limit $\beta_S = 1 / \sqrt{3}$, in units of $c$. Expressed in the lab frame this speed is further reduced by a factor $\gamma$, the Lorentz factor of the flow. The flow is therefore effectively along radial lines initially and independent of $\theta$. Additionally, in the lab frame, all swept-up material is concentrated in an extremely thin shell of width $\Delta R \propto R / \gamma^2$, where $R$ the blast wave radius (here, one $\gamma$ follows from going from comoving to lab frame density, the other from Lorentz contraction of the shell width). This means that initially dimensionless coordinate $A \uparrow 1$ across the entire shell and the fluid equations end up being a function of $B$ only, implying a self-similar solution describing the fluid evolution exists at least to leading order in $1 / \gamma^2$. This is indeed the case and in the Blandford-McKee (BM) solution \cite{Blandford:1976uq} the full fluid profile is known analytically from combining the constraint of self-similarity with conservation of explosion energy within the expanding blast wave.

At very late stages the outflow becomes spherical regardless of its initial collimation and $\theta$ and $\theta_0$ drop out of the equations. The flow becomes non-relativistic as well, $A \downarrow 0$, and again a self-similar solution exists. In the Sedov-Taylor-von Neumann solution (e.g. \cite{Sedov:1959}), the entire fluid profile is again known analytically and the radius of the blast wave can be immediately deduced up to a multiplicative constant just from dimensional analysis, leading to a combination of parameters identical to that for $B$.

A disadvantage of the self-similar solutions is that they do not apply to the intermediate stage of deceleration. However, it is straightforward to construct simplified dynamical models describing the entire evolution for the spherical case once one assumes that all swept-up mass is concentrated in a thin homogeneous shell near the shock front and various such models exist in the literature \cite{Piran1999, Chiang1999, Huang1999, Peer2012}. The common feature of these models is that by combining the shock-jump conditions with energy conservation, a prescription for the evolution of the blast wave Lorentz factor can be found. Since many numerical studies use an equation of state (EOS) relating pressure $p$ to internal energy density $e$ that approximates analytically the exact solution for a (trans-)relativistic ideal gas, it is instructive to demonstrate a shell model for one such EOS,
\begin{equation}
p / (\rho c^2) = \frac{e / (\rho c^2)}{3} \frac{2 + e / (\rho c^2)}{1 + e / (\rho c^2)},
\end{equation}
which was taken from \cite{Mignone2005} and has been applied, for example, in \cite{Zhang2009, vanEerten2010, vanEerten2010transrelativistic, vanEerten2011chromaticbreaks}. Here $\rho$ is comoving density, $e$ does not include rest mass. The correct asymptotic limits are retrieved: $p = e / 3$ and $p = 2e / 3$ in the relativistic and non-relativistic case respectively. For this EOS, the general shock-jump conditions for a blast wave in a cold medium become simply (see also \cite{Uhm2011}):
\begin{equation}
\rho = 4 \gamma \rho_{ext}, \quad e = 4 \gamma(\gamma - 1) \rho_{ext} c^2, \quad p = 4 (\gamma^2 - 1) \rho_{ext} c^2 / 3,
\end{equation}
and tell us, for example, that the jump in density at the shock front $\rho / \gamma$ will be equal to 4 throughout the \emph{entire} evolution of the blast wave (see also \cite{vanEerten2010transrelativistic}). It further follows that the width of the homogeneous shell is \emph{always} $\Delta R = R / (12 \gamma^2)$, if it is to contain all swept-up mass $M$ with density given by the jump condition. The shell volume is then given by $V_S = M / (4 \rho_{ext} \gamma^2)$. The dynamics of the shell follow from fixing the total energy in the shell (here expressed in the lab frame):
\begin{equation}
E_{iso} = \tau V_S = [(\rho c^2 + e + p) \gamma^2 - p - \gamma \rho c^2] M / (4 \rho_{ext} \gamma^2),
\end{equation}
leading to
\begin{equation}
 E_{iso} / (M c^2) = \beta^2 ( 4 \gamma^2 - 1 ) / 3.
 \end{equation}
The ultra-relativistic limit has $\gamma \propto M^{-1/2} \propto t^{-3/2}$, and the non-relativistic limit $\beta \propto M^{-1/2} \propto t^{-3/5}$, as expected from the self-similar solutions. Solving the shell model reveals the enormous range of distance scales involved, which is the key numerical challenge. Simulating the deceleration of a typical BM type blast wave with $E_{iso} = 10^{53}$ erg and $\rho_{ext} \equiv n_{ext} m_p = m_p$ (i.e. one proton cm$^{-3}$) from $\gamma = 100$ onwards until $\beta \gamma \sim 10^{-2}$ means going from $10^{17}$ cm to $10^{20}$ cm, while the initial shell width $\Delta R \sim 10^{14}$ cm. Simulations therefore typically require adaptive-mesh refiniment (AMR, where the grid resolution is dynamically and locally adapted leading to an effective resolution that can be orders of magnitude larger than the base grid resolution). Alternative and complementary approaches exist, such as using moving grid boundaries \cite{Mimica:2008up}, setting up the simulation in a Lorentz boosted frame \cite{vanEerten:2012xk} or using (multi-dimensional) Lagrangian methods where the grid cells advect with the flow \cite{Kobayashi1999, Duffell2011, Duffell2013}.

Arguably the most obvious generalization in terms of dynamics is changing the circumburst density environment to one where density depends on radius as a power law. For long GRB's, where the progenitors are thought to be massive stars \cite{Woosley:1993wj, MacFadyen:1998vz}, one would expect the environment to resemble a stellar wind, $\rho_{ext} \propto r^{-2}$, presumably generated by a Wolf-Rayet type progenitor star \cite{Chevalier:1999mi}.

A number of authors have performed numerical studies of BM type blast waves decelerating in a stellar wind environment \cite{Nakar2007, Meliani2007, DeColle:2011ca}. The effect of an environment $\rho_{ext} = \rho_0 (r/r_0)^{-k}$ is that for higher $k$ the blast wave takes more time to decelerate, and the characteristic time scales change accordingly \cite{Piran2005}. Our shell model, for example, reaches $\beta \gamma = 1$ at $t_{NR} \approx 922 [(E_{iso} / 10^{53}) (m_p r_0^{-k}/ \rho_0) (3-k)]^{1 / (3-k)}$ days. The observational implication is that characteristic features (such as jet breaks, see below) will be stretched out over time. The numerical implication is that a larger grid and longer running time are required to capture the same dynamical stages. The blast wave profiles are scale invariant between energies and densities for each $k$, although the dimensionality of $\rho_0$ needs to be taken into account when expressing scale invariance in terms of $E_{iso}$ and $\rho_0$ \cite{vanEerten:2012xk}.

Further generalizations to jet dynamics include adding structure to the initial outflow (e.g. \cite{Rossi2002, Berger2003}), increasing the mass of the initial ejecta (as included in the original fireball model, see also e.g. \cite{Kobayashi1999, Duffell2013}) or prolonging the duration of energy injection (with initial mass and energy injecting both giving rise to a reverse shock) or taking into account complex circumburst medium structures and transitions (e.g. \cite{Eldridge2006, vanMarle2006, Peer2006, Mesler2012, vanMarle2012, Gat:2013zla}). When the shock dynamics are numerically resolved, it is found to be very difficult to model strong variability in afterglow light curves through circumburst medium interactions \cite{Nakar2007, vanEerten2009opticalvariability, Mimica2011, Gat:2013zla} (although it does offer a plausible explanation for late time shallowing of the light curve, \cite{Gat:2013zla}). The explanation of afterglow flares therefore most likely requires some form of magnetic reconnection (e.g. \cite{Giannios2006}) or late engine activity (see e.g. \cite{Sari2000, Perna2006, Maxham2009, Vlasis:2011yp}).

\section{Emission}

If an expanding relativistic blast wave contains a non-thermal distribution of electrons and when magnetic fields are present, synchrotron emission naturally follows. Detailed theoretical analyses of the standard model basically follow  \cite{Blandford:1977}, where synchrotron and synchrotron-self Compton (SSC) emission are discussed in the context of the self-similar BM solution established previously by the same authors \cite{Blandford:1976uq}, although that article predates the discovery of afterglows by twenty years.


The standard fireball model approach to afterglows (e.g. \cite{Wijers1997, Sari1998, Granot:2001ge}) assumes that a non-thermal distribution of electrons with distribution $n_e (\gamma_e) = C_e \gamma_e^{-p}$ (and $n_e$ and $\gamma_e$ expressed in the frame comoving with the local fluid element) is generated through shock-acceleration at the front of the blast wave. The energy distribution index $p$ (which has nothing to do with pressure $p$) typically lies between 2 and 3, and the distribution cuts off below at $\gamma_m$. If $p \le 2$, the total energy density in accelerated electrons (defined here excluding rest-mass) $\int (\gamma_e - 1) n_e(\gamma_e) m_e c^2 \dev \gamma_e$ diverges if no upper boundary $\gamma_M$ is included. When it is assumed (1) that the energy density in non-thermal electrons is a fraction $\epsilon_e$ of the available internal energy density $e$ and (2) that a fixed fraction $\xi_n$ of the available electrons $n$ are accelerated (with $n$ also the proton number density in the fluid), we can determine $\gamma_m$ and $C_e$. For $\gamma_m$ (and assuming $\gamma_M \uparrow \infty$ at the acceleration site), we obtain:
\begin{equation}
\gamma_m = \frac{p-2}{p-1} \left( \frac{\epsilon_e e}{\xi_n n m_e c^2} + 1 \right).
\label{gamma_m_equation}
\end{equation}
The rest mass term on the right is usually ignored, assuming $\gamma_m \gg 1$. This becomes less accurate as time goes on and $\gamma$ decreases. Even when rest mass is included, we find that the $\gamma_m = 1$ threshold is crossed when $\gamma = 1 + \xi_n m_e m_p^{-1} \epsilon_e^{-1} (p - 2)^{-1}$, at which point the parametrization breaks down. At very late times, therefore, $\xi_n$ \emph{must} be smaller than unity, as is the case for supernova remnants. Alternative parametrizations of the shock-microphysics that deal with late times are possible, see e.g. \cite{Huang2003, vanEerten2010transrelativistic}.

Following shock-acceleration, the population of electrons evolves according to
\begin{equation}
 \frac{\dev \gamma_e}{\dev t} = - \frac{4 \sigma_T \gamma_e^2}{3 \gamma m_e c} (U_B + U_{IC}) + \frac{\beta_e^2 \gamma_e}{3 n} \frac{\dev n}{\dev t},
\label{kinetic_equation}
\end{equation}
assuming that the population remains confined to its fluid element \cite{Downes2002, Granot:2001ge} (an assumption that can be found to be justified up to very high $\gamma_e$ by checking the Larmor radius of the accelerated electrons). The first term on the right contains magnetic field energy density $U_B$, photon field energy density $U_{IC}$ and Thomson cross section $\sigma_T$ and represents energy loss due to synchrotron and synchrotron self-Compton radiation. Energy loss due to adiabatic evolution of a fixed volume in phase space is given by the second term. The electron velocity $\beta_e$ can safely be assumed to be 1 for a relativistic population. It is only electrons with $\gamma_e \gg 1$ that emit synchrotron radiation.

Once the flow becomes non-relativistic, $e \propto n^{5/3}$ rather than $e \propto n^{4/3}$, implying that once $\epsilon_e$ is set to some parametrized value (typically around 0.1) at the shock front, adiabatic evolution of the relativistic electron population will cause it to evolve further downstream according to $\epsilon_e \propto e^{-1/5}$. Since energy density decreases downstream, the relative energy content of the relativistic electrons will grow. Once radiative losses are accounted for as well, the evolution becomes even more complex. Because the adiabatic evolution induced dependency of $\epsilon_e$ on $e$ is only weak and because the evolution of $\gamma_m$ is typically dictated by the adiabatic loss term only, it is usually assumed in numerical studies for the purpose of determining $\gamma_m$ that $\epsilon_e$ remains fixed as a fluid element advects downstream post-shock \cite{Nakar2007, Mimica:2008up, Zhang2009, vanEerten2010transrelativistic, vanEerten2011chromaticbreaks, Wygoda:2011vu, DeColle:2011ca, Mimica2011}.

Each individual electron emits a synchrotron spectrum peaking at
\begin{equation}
\nu'_e(\gamma_e) = \frac{3 q_e}{4 \pi m_e c} \gamma_e^2 B,
\end{equation}
measured in the frame comoving with the fluid element, with $B$ magnetic field strength and $q_e$ electron charge. The flux from an individual electron will drop exponentially at higher frequencies. In the absence of electron cooling, the local accelerated electron population as a whole will emit a synchrotron spectrum peaking at $\nu'_m \propto \gamma_m^2 B$ and with emission coefficients $j_\nu$ asymptoting to $j_\nu \propto (\nu' / \nu_m')^{1/3}$ below and $j_\nu \propto (\nu' / \nu_m')^{(1-p)/2}$ above $\nu'_m$, with the exponential cut-offs of individual electrons adding up to a power law slope. The observed global synchrotron spectrum consists of the combined emission of all regions in the blast wave and will have the same asymptotic shape. The exact shape of the local and global spectrum and the flux level at the peak depend on the amount of detail used in modeling synchrotron emission and people use both simply connected power laws (e.g. \cite{Wijers1997, Sari1998, Zhang2009}) or detailed expressions with smooth spectral transitions based on full integration of modified Bessel functions (e.g. \cite{Granot:2001ge, vanEerten2009, Leventis2012}).

Typically, the magnetic field required for synchrotron emission is assumed to be small scale, randomly oriented and generated at the shock front. The magnetic energy density $U_B$ is parametrized by linking it to the internal energy density according to $U_B \equiv B^2 / (8 \pi) \equiv \epsilon_B e$, with $\epsilon_B$ typically of the order 0.01. This results in magnetic fields of strength $B \sim 0.6 (\epsilon_B / 0.01)^{1/2} (n_{ext} / 1. $ cm$^{-3}) (\gamma / 10.)^2$ Gauss for relativistic blast waves, much larger than what can be obtained by shock-compression by a factor of $4 \gamma$ \cite{Gallant1999, Achterberg2001} of an ambient circumburst magnetic field with field strength on the order of $\mathrm{\mu G}$. In most cases, a shock-compressed ambient field is insufficient to explain the data, but some interesting exceptions exist \cite{Kumar2010, BarniolDuran2011}. 

Ultimately, due to the complexity of particle acceleration and magnetic field generation at shocks, massive numerical computations of large groups of individual particles accelerating and interacting (``particle-in-cell'' or PIC simulations) are important to obtain a physical understanding of the magnetic fields and non-thermal populations underlying gamma-ray burst emission (e.g. \cite{Spitkovsky2008, Sironi2009}). These computations can then in principle be used to inform macrophysical parametrizations (e.g. $\epsilon_e$, $\epsilon_B$, $p$, $\xi_n$) and, in that way, eventually be compared to observational data. At the moment the spatial and temporal scales covered by the particle-in-cell simulations are unfortunately still limited by computer power and no convergence has been achieved for the emergent properties of the system.

\el

Strictly speaking, a power law distribution of particles injected at the shock front will not remain a power law distribution further downstream, mainly through the effect of radiative cooling. Even when $1 / \gamma_M$ initially starts out near zero, it will evolve quickly according to eq. \ref{kinetic_equation}, leading to an exponential drop in flux for a local particle population beyond $\nu'_M \equiv \nu'_e(\gamma_M)$. The cut-off $\gamma_M$ can in principle be determined by comparing the acceleration time scale to the radiative cooling scale and will typically lead to $\nu'_M$ of the order GeV (e.g. \cite{Norman1995, Blandford1987, Peer2013}). Numerically and in simplified analytical models, cooling is often dealt with by assuming that for the purpose of calculating electron cooling effects, a global steady state exists in the shocked plasma where above a certain electron $\gamma_c$ the radiative loss term and energy injection term due to shock acceleration are in equilibrium. This then leads to a steepening of the spectrum by $1/2$ beyond $\nu'_c \equiv \nu'_e(\gamma_c)$ when $\nu'_c > \nu'_m$ (``slow cooling'') and, in case $\nu'_c < \nu'_m$ (``fast cooling''), a spectral slope transition from $1/3$ to $-1/2$ across $\nu'_c$ and eventually to $-p/2$ beyond $\nu'_m$. The power law steepening is consistent with the emergent spectrum for local cooling, where all the exponential drops at locally different cut-off frequencies add up to a power law. The cooling break Lorentz factor $\gamma_c$ is obtained from a rough estimate where the cooling time is equated to the life time of the blast wave, leading to
\begin{equation}
\gamma_c = 6 \pi m_e c \gamma / (\sigma_T B^2 t).
\label{global_cooling_time_equation}
\end{equation}

An alternative approach would be to solve eq. \ref{kinetic_equation} along with the hydrodynamic equations. This was done analytically for the BM solution in \cite{Granot:2001ge}, and numerically in e.g. \cite{Downes2002, Nakar2007, vanEerten2010transrelativistic, Uhm2013}. Locally calculated cooling will impact both the overall flux level and the sharpness of the cooling break \cite{Granot:2001ge, vanEerten2010, Uhm2013}. The resolution required to solve the cooling locally follows from considering
\begin{equation}
\Delta (R-r) \approx \Delta \gamma_e^{-1} \frac{\dev (R-r)}{\dev t} \frac{3 \gamma m_e c}{4 \sigma_T U_B},
\label{emission_region_size_equation}
\end{equation}
which can be derived from eq. \ref{kinetic_equation}, assuming the fluid conditions don't change across the hot electron region $\Delta (R-r)$. We want to resolve $\Delta \gamma_e^{-1}$ going from 0 to, say, the Lorentz factor associated with emission peaking at X-rays ($\nu \sim 5 \times 10^{17}$ Hz). When this is done for the shell model described previously and for typical afterglow values ($E_{iso} = 10^{53}$ erg, $n_{ext} = 1$ cm$^{-3}$, $\epsilon_B = 0.01$, $\epsilon_e = 0.1$), it is found that $\Delta (R-r) / \Delta R$ starts around 0.5 when $\gamma = 100$, decreases with $\Delta (R-r) / \Delta R \propto \nu^{-1/2} \gamma^{2/3}$ as the blast wave decelerates, plateaus at $\sim 5 \times 10^{-2}$ around $\beta \gamma \sim 1$ before decreasing again according to $\Delta (R-r) / \Delta R \propto \nu^{-1/2} \beta^{1/6}$. What this means is that although the size of the hot region is comparable to the blast wave width at high Lorentz factors, thus allowing for approximations like eq. \ref{global_cooling_time_equation}, this approximation gets progressively less accurate as the blast wave decelerates. It also means that it is very challenging to numerically model local cooling by rewriting eq. \ref{kinetic_equation} into an advection equation, given that the resolution requirement increases by a factor $\Delta R / \Delta (R-r)$ (but not impossible, see \cite{Downes2002, Nakar2007, vanEerten2010transrelativistic}; a Lagrangian approach is recommended in order to accurately detect the position of the shock front).

Note that, analytically, the same light curve power law behavior follows whether it is derived assuming a finite emission region (i.e. eq. \ref{emission_region_size_equation}) plus sharp emission cut-off or assuming emission from the full blast wave plus power law change in spectrum: the differences between the two approaches will mostly be apparent during sudden transitions in the outflow, such as a jet break or the rise of a reverse shock in the case of massive ejecta or change in the nature or the circumburst medium.

With a quantative model for the synchrotron emission including cooling, it is possible to check one key assumption mentioned previously: that of adiabatic expansion. A calculation of the total emitted power in synchrotron emission for our shell model, global cooling and typical afterglow values, reveals this to be a safe assumption. When the blast wave $\beta \gamma$ drops to $10^{-2}$, the total energy loss is found to be about 2 percent of $E_{iso}$.

\el

At low (typically radio) frequencies, the blast wave becomes opaque due to synchrotron self-absorption (SSA). Like the synchrotron emission coefficient $j_\nu$, the SSA coefficient $a_\nu$ can be modeled at varying levels of detail. Analytical scaling models often simply consider the asymptotic limit where the emitting volume is replaced by an emitting outer surface \cite{Sari1998, Waxman1998}. Alternatively, an implementation of linear radiative transfer can be used \cite{vanEerten2010transrelativistic, Mimica2010, vanEerten2010, vanEerten2011chromaticbreaks, vanEerten2010} with either a simple power law approximation to $a_\nu$ or a more complete treatment with smooth transitions \cite{Leventis2012}, which can even include the effect of electron cooling on $a_\nu$ \cite{Granot:2001ge}.

A useful property of synchrotron spectra is that they too exhibit scale invariance between energies and between circumburst densities in their asymptotic spectral regimes \cite{vanEerten2012scaleinvariance}, even when computed numerically from two dimensional simulations of spreading trans-relativistic blast waves. Although perhaps less obvious than the invariance in dynamics, this invariance amounts to the same thing and works because in the different power law regimes, additional constants with dimension entering into the flux formulae (i.e. $m_p$, $m_e$, $\sigma_T$) can be identified and grouped together, leaving the remaining terms again fixed by dimensional analysis.

\el

It is not difficult to come up with physically plausible complications to the standard synchrotron radiation model. Synchrotron Self Compton (SSC) was already briefly mentioned above and can be expected to impact afterglows at gamma-rays and hard X-rays, especially for high blast wave Lorentz factors \cite{Sari2001, Petropoulou2009}. I already mentioned pitfalls of using $\xi_N$ and $\epsilon_e$. In addition, the downstream evolution of $\epsilon_B$ depends on the nature of the magnetic field. A randomly oriented field with fixed number of field lines through the surface of each fluid element will evolve such that $e_B \propto \rho^{4/3}$, meaning that $\epsilon_B$ remains fixed only for relativistic flows \cite{vanEerten2010transrelativistic}. A preferred direction for the field will further complicate matters, and is an issue best addressed through polarization measurements (reviewed elsewhere in these proceedings). Finally, of the standard radiation parameters, the behavior of $p$ is likely to be more complex than usually assumed. Model fits to various afterglow datasets yield a distribution of $p$ values arguably inconsistent with a single underlying universal value \cite{Curran2010}. This would indicate that $p$ is sensitive to the physical conditions at the front of the blast wave and one would then naturally expect $p$ to evolve strongly over time within each burst as well, since the conditions across the blast wave shock front change across a wide range during the evolution of each blast wave. Although there is theoretical support for $p$ evolution across the transrelativistic regime \cite{Keshet2005}, the sample studied in \cite{Curran2010} is mostly relativistic. Generally, Swift burst data shows no clear temporal trends or variability for the spectral index, although the error bars are usually large.

\section{The jet nature of the outflow}

At some point the outflow will no longer follow radial lines but bend sideways, revealing the collimated nature of the blast wave. For a blast wave initially following the BM solution, the Lorentz factor of the shock $\Gamma$, can be found to obey
\begin{equation}
\Gamma^2 = (17 - 4k) E_{iso} / (8 \pi \rho_{ref} R_{ref}^k t^{3-k} c^{5-k}),
\end{equation}
with the numerical constants following from radial integration over the BM lab frame energy density profile. A sound wave traveling along the shock front will have its radial component in the lab frame set by $\Gamma$ in order to keep up with the outward motion of the shock. Since its magnitude in the comoving frame is also known (i.e. $\beta_S = 1 / \sqrt{3}$), the transverse component of the sound speed in the lab frame can be calculated to be $\beta_{S,\theta}' = 1 / 2 \Gamma$. Integrating $R \dot{\theta} = \beta_{S,\theta}'$ for a sound wave traveling at $t=0$ from the jet edge at $\theta_0$ to the tip then yields $\Gamma_j = (3 - k)^{-1} \theta_0^{-1}$ for the shock Lorentz factor at which the tip and the blast wave as a whole begin to decelerate and a qualitative change in the nature of the flow sets in.

In the limiting case of ultra-narrow and ultra-relativistic jets, this new stage can be shown analytically to be one where the Lorentz factor drops exponentially, while the opening angle $\theta_{max}$ widens exponentially \cite{Rhoads:1999wm, Gruzinov:2007ma} once $\theta_{max} \gg \theta_0$.  This follows from the fact that a widening jet sweeps up more mass, leading to further deceleration which increases sideways expansion in the lab frame etc., leading to a runaway effect. In practice, this regime is not found to occur for jets with typical opening angles ($\theta_0 \sim 0.1$ rad, \cite{Frail2001}), since by the time $\theta_{max} \gg \theta_0$ the jet is no longer in the ultra-relativistic regime. Note that e.g. for $\theta_0 \sim 0.05$, the fluid Lorentz factor of the tip $\gamma \approx 4.7$ at the \emph{onset} of deceleration, leaving no room for an intermediate stage where $\gamma \ll 1/ (3 \theta_0 \sqrt{2} )$ \emph{and} $\gamma \gg 1$ \emph{and} $\theta \gg \theta_0$.

The slow expansion in practice of afterglow jets has been confirmed numerically by various authors \cite{Granot:2001cw, Zhang2009, vanEerten2010, vanEerten2011chromaticbreaks, Wygoda:2011vu, DeColle:2011ca, vanEerten2012observationalimplications}. A phase of exponential expansion can be found \cite{MacFadyen2013} for jets with $\theta_0 \ll 0.04$, but these simulations require a special approach such as a boosted frame due to the requirement $\gamma \gg 1 / \theta_0$ for the initial conditions discussed previously. Advanced analytical models incorporate a smooth transition between the exponentional and logarithmic stages of spreading \cite{Wygoda:2011vu, Granot:2011gg} (see also \cite{vanEerten2012observationalimplications, MacFadyen2013}).

The jet nature of the blast wave will become apparent to an observer in two ways, both leading to a steepening of the light curve. On the one hand, due to strong relativistic beaming, an observer originally only sees a small patch of the blast wave surface. Once the blast wave has decelerated sufficiently, and the relativistic beaming cones (with width $\theta \sim 1 / \gamma$) have widened sufficiently, this patch will have grown to include the edge of the blast wave and a lack of emission from beyond the edges will cause the observed flux to decrease more steeply. On the other hand, the decrease in beaming due to the additional deceleration caused by the spreading of the jet will also lead to a steeper decrease of the observed flux. Since jet spreading is not as extreme as originally thought, both effects contribute noticeably and the first effect is not overwhelmed by the second. A specific consequence of this is that the shape and onset of the jet break become different even for small changes in observer angle, even when still within $\theta_0$. As opposed to the second, dynamical, cause for the jet break, the onset and completion of the first effect depend on the angle between observer and outer edges (i.e. on $\theta_0 \pm \theta_{obs}$) rather than on $\theta_0$ alone. This implies that jet breaks are stretched out over time and often do not become fully clear until sufficient time has passed, which might be beyond the capabilities of Swift to observe. This provides a natural explanation \cite{vanEerten2010} for the lack of clear jet breaks detected by Swift \cite{Kocevski2008, Racusin2009}.

\section{Comparison to data}

Ultimately, we wish to compare between model and data. A number of approaches are possible. One can fit basic functions, such as (smoothly broken) power laws to the various bands and subsequently interpret these. Or one can directly fit analytical or simulation-derived synthetic light curves. The latter has the advantage of potentially getting the most out of the data, but the number of free parameters of the standard afterglow model ($E_{iso}$, $\theta_0$, $\rho_0$, $p$, $\epsilon_e$, $\epsilon_B$, $\xi_N$, observer angle $\theta_{obs}$) can be problematic. Full broadband afterglow datasets covering the full range from radio to X-rays (and thus all spectrum regimes) are very rare. Solutions are to either add constraints to the model (e.g. $\epsilon_B \equiv \epsilon_e$ or $\xi_N \equiv 1$) or carefully study the probability distributions of the various fit parameters in order to determine what is and what isn't constrained. A Bayesian approach is well suited to this task (see e.g. the contributions by B.B. Zhang and by Ryan elsewhere in these proceedings), and would, for example, naturally bring out the extent to which the degeneracy between $\xi_N$ and other model parameters \cite{Eichler2005, Leventis2012} is broken by the strict upper limit of 1 on $\xi_N$.

\Acknowledgements

This research was supported in part by NASA through grant NNX10AF62G issued
through the Astrophysics Theory Program and by the NSF through grant AST-
1009863 and by the Chandra grant TM3-14005X.


\end{document}